\documentclass[preprint,review,5p,times]{elsarticle}

\usepackage{lineno,hyperref}
\modulolinenumbers[5]
\usepackage[labelfont=bf]{caption}
\journal{Journal of \LaTeX\ Templates}
\usepackage[utf8x]{inputenc}
\usepackage{hyperref}
\usepackage{amssymb}
\usepackage{enumitem}
\setlist{nosep}
\usepackage{graphicx}

\begin{document}

\begin{frontmatter}

\title{A facility for direct measurements for nuclear astrophysics at IFIN-HH - a 3 MV tandem accelerator and an ultra-low background laboratory}

\author[rvt,focal]{D. Tudor}

\author[rvt]{L. Trache\corref{cor1}}
\ead{livius.trache@nipne.ro}

\author[rvt,focal]{A.I. Chilug}
\author[rvt,focal]{I.C. Stefanescu}
\author[rvt]{A. Spiridon}
\author[rvt]{M. Straticiuc}
\author[rvt]{I. Burducea}
\author[rvt]{ A. Pantelica}
\author[rvt]{R. Margineanu}
\author[rvt]{ D.G. Ghița}
\author[rvt]{D. G. Pacesila}
\author[rvt]{R. F. Andrei}
\author[rvt]{C. Gomoiu} 

\author[els]{ N.T. Zhang}
\author[els,add4]{ X.D. Tang}

\cortext[cor1]{Corresponding author}

\address[rvt]{Horia Hulubei National Institute for R$\&$D in Physics and Nuclear Engineering, IFIN-HH, 077125 București-Măgurele, Romania}
\address[focal]{Doctoral School of Physics, University of Bucharest, 077125 București-Măgurele, Romania}
\address[els]{Institute of Modern Physics, Chinese Academy of Sciences, Lanzhou 730000, China}
\address[add4]{Joint department for nuclear physics, Lanzhou University and Institute of Modern Physics,Chinese Academy of Sciences, Lanzhou 730000, China}

\begin{abstract}
We present a facility for direct measurements at low and very low energies typical for nuclear astrophysics. The facility consists of a small and robust tandem accelerator where irradiations are made and an ultra-low background laboratory located in a salt mine where very low radio-activities can be measured. Both belong to “Horia Hulubei” National Institute for Physics and Nuclear Engineering (IFIN-HH), but are situated 120 km apart. Their performances are shown using a few cases where they were used. We argue that this facility is competitive for the study of nuclear reactions induced by alpha particles and by light ions at energies close to or down into the Gamow windows. A good case study was the $^{13}$C+$^{12}$C fusion reaction, where the proton evaporation channel leads to an activity with T$_{1/2}$=15 h, appropriate for samples' transfer to the salt mine. Measurements were done using the thick target method down into the Gamow window for energies from E$_{c.m.}$= 2.2 MeV, which is the lowest energy ever reached for this reaction, up to 5.3 MeV, using $^{13}$C beams from the 3 MV Tandetron. The activation method allowed us to determine a cross section of the order of 100 pb. Reactions induced by alphas were also measured. Proton induced resonant reactions were used to calibrate the accelerator terminal voltage. Some results of the experiments characterizing the assembly are shown and discussed.
\end{abstract}

\begin{keyword}
\texttt{nuclear physics for astrophysics, direct measurements, thick target method, activation methods,  ultra-low background}
\end{keyword}

\end{frontmatter}

\section{ Introduction}

\par{Nuclear astrophysics (NA) is for some time already an important part of the science program of most nuclear physics laboratories. The experimental studies can be divided as direct measurements – reactions studied at the low energies as they happen in stars, or as close to that as possible, followed by extrapolations into the so called Gamow window – and indirect methods, where information (nuclear data) is extracted from reactions at much larger energies, information that is then used to evaluate the reaction cross sections or the reaction rates in the region of energies relevant for astrophysics. This is due to the fact that at low energies the reactions involving charged particles – and this is a large part of reactions in stellar environments – are very much hindered by the Coulomb barrier, leading to considerable measurement difficulties \cite{1988R}. Therefore, the case of direct measurements calls for special experimental solutions. One of them is to install high intensity accelerators in underground laboratories. The first such and best known is the LUNA project \cite{LUNA} at the Laboratori Nazionali di Gran Sasso of INFN, in Gran Sasso, Italy. Several other projects are under development or in planning phase in USA and China. To install an underground facility is not an easy task and, therefore, dedicated projects for nuclear astrophysics could so far be planned around existing or planned larger underground physics laboratories. 
We present here the case where we combine the use of a new small accelerator situated at the surface on the premises of the IFIN-HH institute with an ultra-low background laboratory the institute has in a salt mine in Slanic-Prahova, about 120 km North of Bucharest. The 3 MV tandem accelerator \cite{3MV} can deliver low energies and relatively large beam current (tens of $\mu$A) of most stable elements from protons up. In the microBequerel ultra-low background \cite{SaltMine} laboratory we can measure then samples with activities down to mBq. That is due to its special natural conditions that lead to very low gamma-ray background from natural radioactivity. If the reactions under study produce activations with life times that allow the transfer to Slanic, we can gain significantly in detection sensitivity.} 
\par{The paper is divided as follows: Sect. 2 describes summarily the accelerator, its beam currents, stability and energy calibration. Section 3 describes the characteristics of the microBequerel laboratory that are important in these particular nuclear astrophysics experiments. The ultra-low gamma-ray background is demonstrated and the HPGe detector efficiency calibration is emphasized. In Sect. 4 we present the main features of the combination accelerator - salt mine laboratory using the $^{13}$C+$^{12}$C reaction at low energies, down into the equivalent of the Gamow window of the $^{12}$C+$^{12}$C reaction of crucial importance for nuclear astrophysics. Results of two other reactions induced by alpha particles are mentioned. Sect. 5 presents the conclusions of the study. Short descriptions of the facilities were included before in preliminary reports of the reactions studied \cite{SantaTecla, IMP, CSSP16} and the results of the $^{13}$C+$^{12}$C fusion reaction studies are submitted for publication elsewhere \cite{IMP1}.}

\section{The 3 MV TANDETRON$^{TM}$ accelerator}

\par
\begin{figure}[h]
	\begin{center}
		\includegraphics[width=0.5\textwidth]{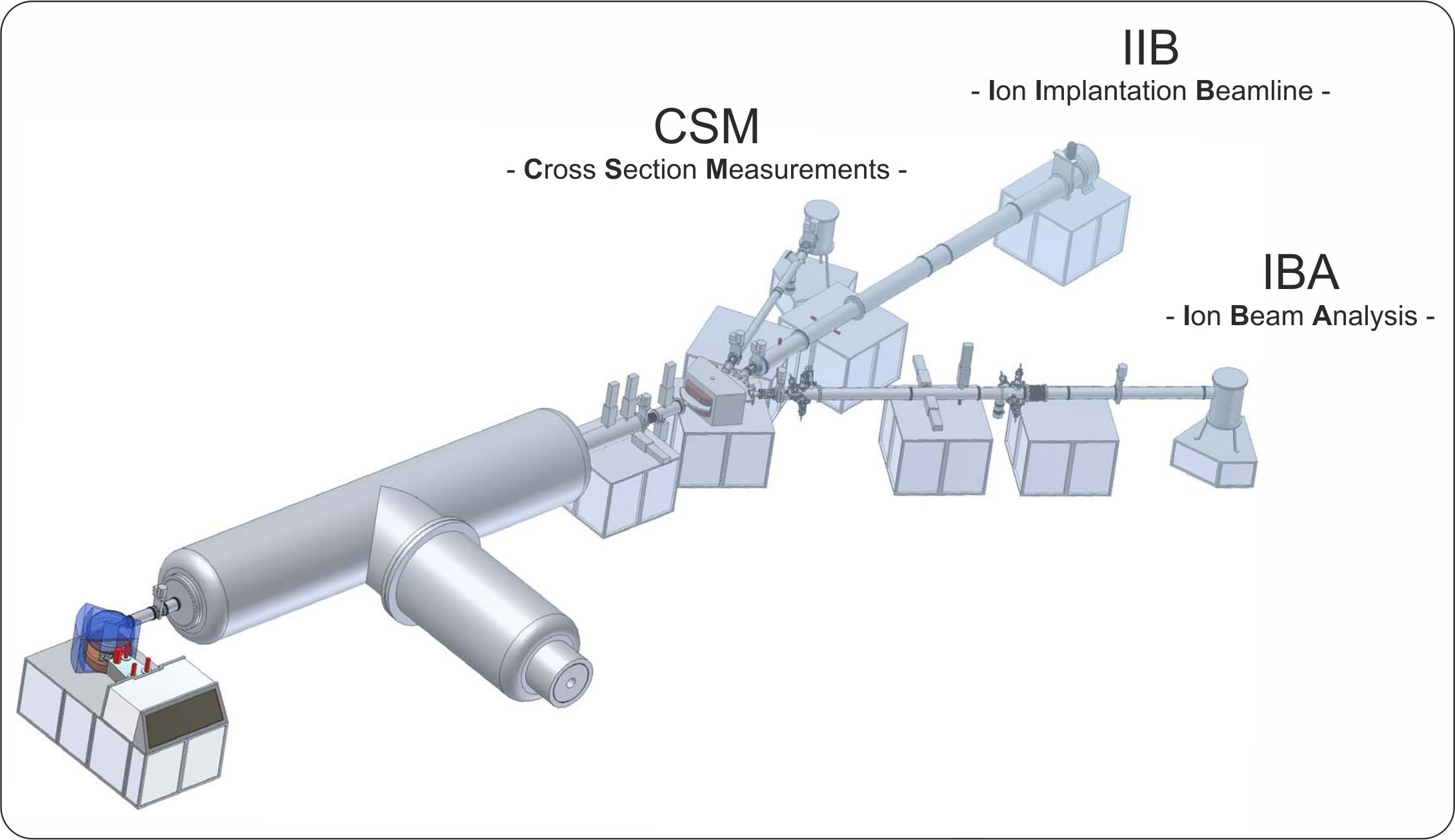}
	\end{center}
	\setlength\abovecaptionskip{-2pt}
	\caption{Outline of the 3 MV tandem accelerator facility.}\label{1}
\end{figure}

\begin{figure}[h]
	\begin{center}
		\includegraphics[width=0.48\textwidth]{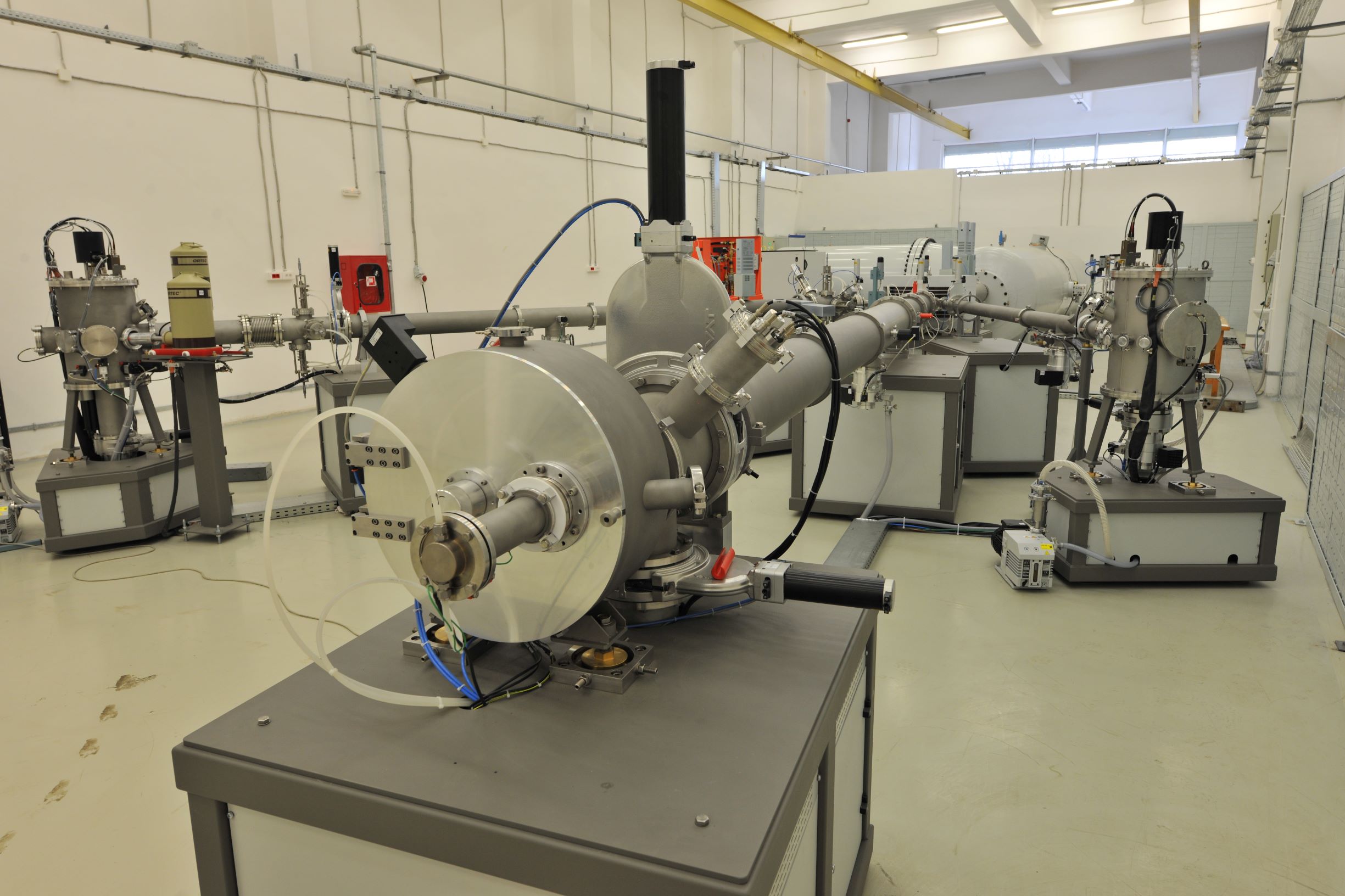}
	\end{center}
	\setlength\abovecaptionskip{-2pt}
	\caption{The 3 MV Tandetron Accelerator at IFIN-HH. NA measurements were done on the beam lines in center and at right.}\label{2}
\end{figure}

{The 3 MV Tandetron$^{TM}$ was designed and built by High Voltage Engineering Europa B.V.  and commissioned at IFIN-HH in 2012. Its original intended use was Ion Beam Analysis (IBA) with various methods: Rutherford Backscattering (RBS), Elastic Recoil Detection Analysis (ERDA), Particle Induced X-rays Emission (PIXE), Particle Induced Gamma-rays Emission (PIGE), Nuclear Reaction Analyses (NRA) and ion implantation, as described in Ref. \cite{3MV}. The layout of the Tandetron$^{TM}$ and its beam lines are shown in Fig. \ref{1}. Fig. \ref{2} is a picture of the overall arrangement in the accelerator hall.}
\par{With the final goal of establishing a solid line of research in nuclear astrophysics at the Bucharest accelerators and laboratories of IFIN-HH, we have performed experiments to check the limits of the one method that seemed  appropriate and for which the institute has or could acquire installations: the activation method. We used for irradiation the new 3 MV Tandetron$^{TM}$ accelerator.}

\begin{table}[h!]
	\begin{center}
		\caption{ Beam currents measured after the 90$^{0}$ deflecting magnet in the accelerator Faraday cup \cite{3MV}. A full list is available on http://www.nipne.ro/research/departments/dfn.php.}\label{17}
		\label{tab:table1}
		\begin{tabular}{l|c|r} 
			Ion source & Ion species & Typical current [$\mu$A]\\
			\hline \hline
			Duoplasmatron & $^{1}$H$^{-}$  & \textgreater40\\ 
			      & $^{4}$He$^{-}$ & \textgreater3 \\        
			\hline
			Cs sputter & $^{11}$B$^{-}$  & \textgreater40\\
			& $^{12}$C$^{-}$  & \textgreater80\\
			& $^{16}$O$^{-}$  & \textgreater80\\
			& $^{28}$Si$^{-}$  & \textgreater80\\
			& $^{31}$P$^{-}$  & \textgreater40\\
			& $^{58}$Ni$^{-}$  & \textgreater70\\
			& $^{63}$Cu$^{-}$  & \textgreater70\\
			& $^{75}$As$^{-}$  & \textgreater10\\
			& $^{197}$Au$^{-}$  & \textgreater80\\
		\end{tabular}
	\end{center}
\end{table}
 We noticed that while there are many small proton accelerators used specifically for NA, some underground, not many accelerators for alpha and light ions are dedicated to nuclear astrophysics direct measurements. This could be a niche. Early on we realized that the accelerator has potential for use in nuclear astrophysics and tested its characteristics. We tested that it:	
\begin{itemize}[label={--}]
	\item can reliably and stably perform for terminal voltages of 0.2 – 3.3 MV;
	\item provides stable currents for long periods of time, typically needed for measurements that need to last days or weeks;
	\item 	provides relatively high currents for a variety of beams. 
\end{itemize}
The beam currents obtained for a few ion species are shown in Table \ref{17}.
In particular, the beam intensities in the order of 1 p$\mu$A for alphas and at least ten times more for $^{12}$C gave us the idea that one can use the accelerator to study light ion-ion reactions of relevance for nuclear astrophysics.

The accelerator high voltage is monitored by a generating voltmeter (GVM) that provides feedback for the Tandetron$^{TM}$ driver. GVM requires periodic calibration and for this work the resonant reaction $^{27}$Al(p,$\gamma$)$^{28}$Si was used \cite{GVM,Calib}. 
\begin{figure}[h]
	\begin{center}
		\includegraphics[width=0.48\textwidth]{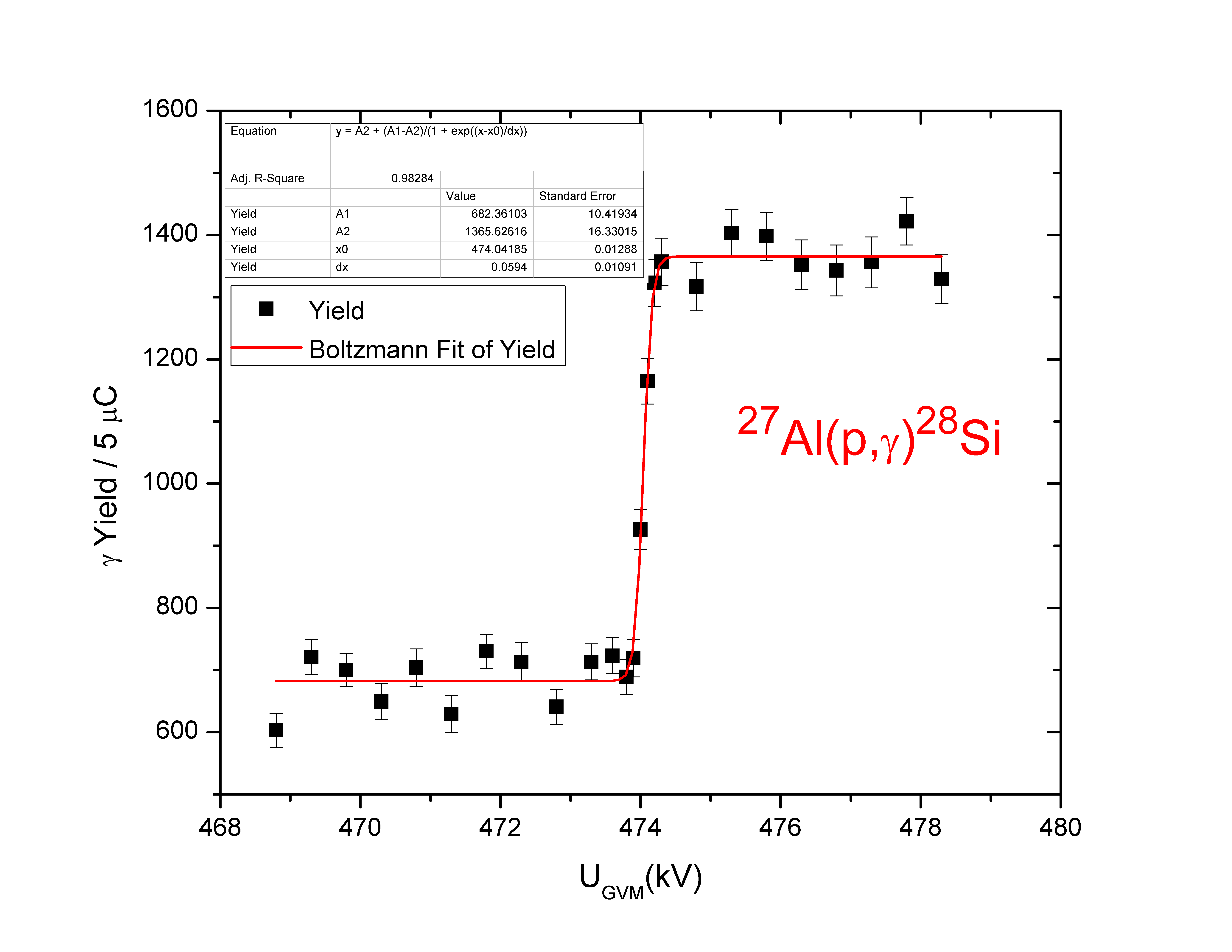}
	\end{center}
	\setlength\abovecaptionskip{-2pt}
	\caption{Excitation function for the $^{27}$Al(p,$\gamma$)$^{28}$Si at 992 keV resonance.}\label{3}
\end{figure}
The well-known narrow resonance at E$_{p}$=992 keV  was scanned in 0.1 kV steps and the excitation function is given in Figure \ref{3}, which is included only to illustrate the actual performances of the procedure. In order to determine the calibration curve two more cross-section maxima were measured near 1317 keV, respectively 1381 keV \cite{Res}. 

\section{The ultra-low background microBequerel laboratory}

\par{Salt mining has in Romania a history that goes back to ancient times, but in Slanic-Prahova the first mine was only opened in 1688. Slanic Prahova is situated about 100 km north of Bucharest. The Unirea salt mine has been open since 1943 with salt exploitation performed until 1970 \cite{Unirea}. After the latter, sections of the mine were opened for visitors. In 2006 the microBequerel (${\mu}$Bq) laboratory of IFIN-HH has been constructed and fully commissioned.
The depth of the mine is around 210 m (${\sim}$600 meter water equivalent) \cite{mwe}. The consideration for which this location has been chosen is the very low natural radioactivity, due to the fact that walls do not present cracks and due to the high purity of the salt \cite{131I}.}
\begin{figure}[h]
	\begin{center}
		\includegraphics[width=0.48\textwidth]{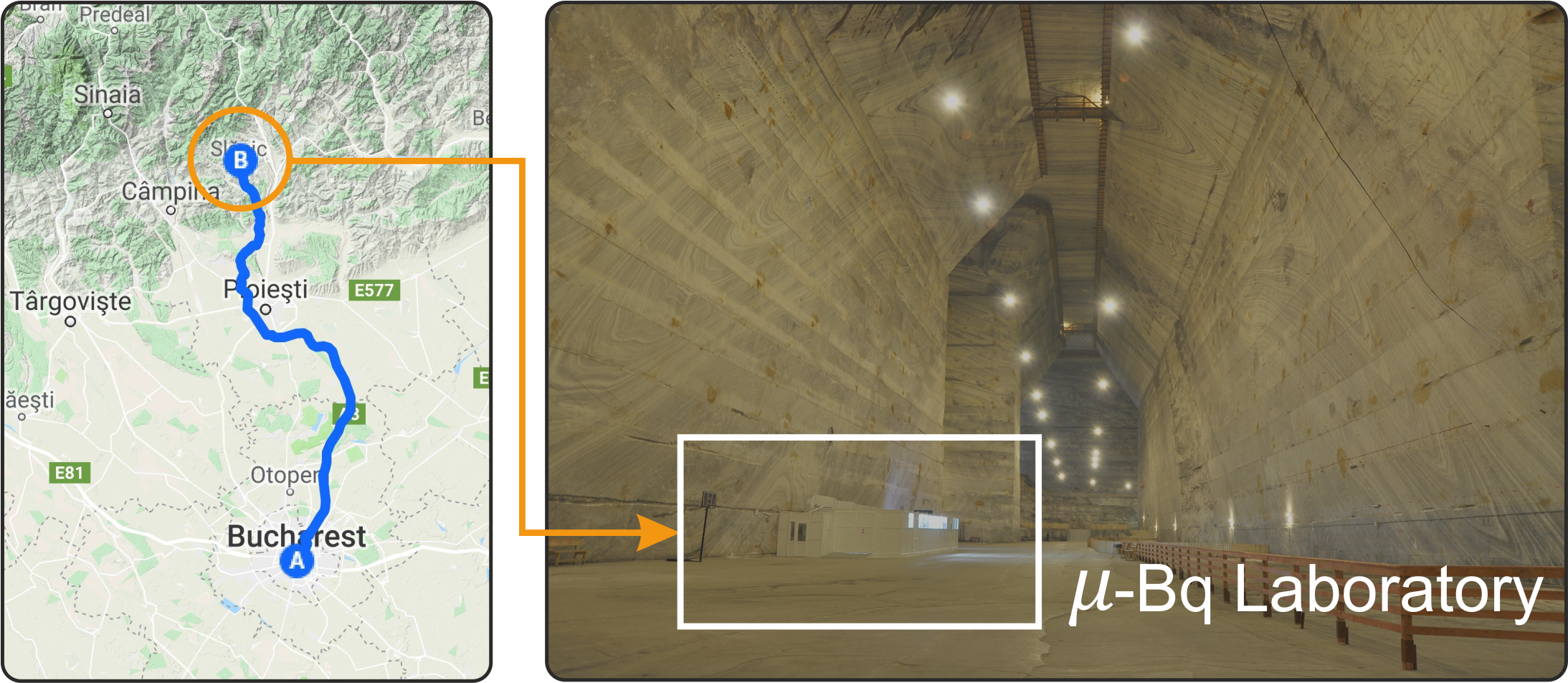}
	\end{center}
	\setlength\abovecaptionskip{-2pt}
	\caption{The location of the mBq laboratory inside the Slanic salt mine.}\label{5}
\end{figure}\\
The Underground Laboratory in the Unirea salt mine, Slanic Prahova (µBq), is located at about 2 hours drive North of Bucharest. Environmental conditions in the salt mine are very stable year-round: temperature between 12 and 13 $^{0}$C, humidity 60-65\% approximately, area of ~70000 m$^{2}$, height between 54 and 58 m, the distance between the walls is between 32 and 36 m, volume is 2.9x10$^{6}$ m$^{3}$. In this mine a laboratory was built and it performs measurements using gamma ray spectrometry in ultra-low radiation background. The average dose underground was found 1.17 ± 0.14 nGy/h, approximately 80-90 times lower than the dose at the surface \cite{SaltMine,131I}. Ambient background radiation comes from: 
\begin{enumerate}[label=\roman*)]
	\item natural radioactivity (especially from the decay of $^{238}$U, $^{232}$Th and $^{40}$K);  
	\item neutrons from ($\alpha$, n) reactions and fission;
	\item cosmic rays ($\mu$, $^{1}$H, $^{3}$H; $^{7}$Be, $^{14}$C ...).
\end{enumerate}

\begin{figure}[h]
	\begin{center}
		\includegraphics[width=0.48\textwidth]{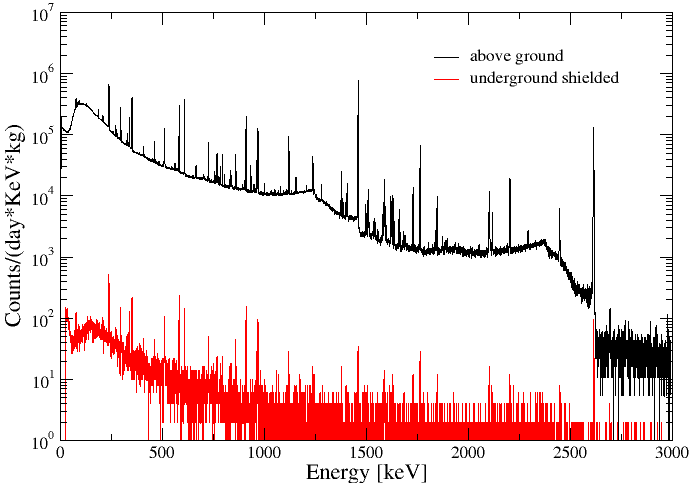}
	\end{center}
	\setlength\abovecaptionskip{-2pt}
	\caption{Natural background from the µBq laboratory collected with the same HPGe: top is above ground, bottom is underground shielded.}\label{6}
\end{figure}
The first two sources are particularly low in this mine due to its thick and compact salt walls. Figure \ref{6} compares ${\gamma}$-ray spectra measured above ground and underground. The top spectrum shows that the strongest components of the $\gamma$ rays spectrum at E$_{\gamma}$\textless2.6 MeV is associated with the natural environmental radioactivity and exhibits intense characteristic lines. At higher energies, the background originates mostly from cosmic rays (iii). The natural radioactivity is significantly reduced for measurements in the underground laboratory (bottom spectrum). From Fig. \ref{6} it can be seen that the measured background radiation (using a protection shield, produced by Canberra Ind., consisting of 15 cm Pb and 5 cm Cu) is about 4000 times smaller compared to the background spectrum measured at the surface. This is the major advantage we wanted to test and use in the current measurements. The total counts from 40 to 2700 keV are compared above. The integrated underground rate for this gamma-ray energy region was 25,870 counts in 48 hours, that is 539(4) counts/hr, (statistical uncertainty only). 
For comparison we can refer  the reader to two underground installations: LUNA at Laboratori Nazionali Gran Sasso \cite{LUNA1} and CASPAR at Sanford Underground Research Laboratory in Lead, SD, USA \cite{CASPAR}. Both consist of accelerators and detection setups, and are very deep under (3800 mwe and 4300 mwe, respectively). Therefore, we can compare only the gamma-ray backgrounds at these places with the one in Slanic, when similar data exist. 
The underground LUNA  facility (accelerator and detectors), under 1.4 km of rock in Gran Sasso, reports  \cite{LUNA1} a rate of 4870 cts/hr in the 1461 keV peak ($^{40}$K) and 1325 cts/hr at the 2614 keV peak ($^{232}$Th series) with a 137 \% relative efficiency HPGe detector. In a similar detector in Slanic we measure a rate of 1.81 cts/hr and 4.8 cts/hr in the same peaks. With special shielding, including anti-radon box with dry nitrogen gas flow around the detector, the rates at LUNA become 0.93 cts/hr and 0.42 cts/hr (setup B in Ref. \cite{LUNA1}) for the same two representative gamma-ray background peaks and with extra shielding these rates were reduced by another factor of 2. More recently, at the location of LUNA2 \cite{LUNA2} the rates for the same two gamma lines are reported as 2190(10) cts/hr and 680(15) cts/hr unshielded, and 14.8(3) cts/hr and 15.2(3), respectively, for the shielded HPGe detector of 100 \% relative efficiency (no anti-radon box). At CASPAR, the gamma-ray background in the region 40-2700 keV is essentially the same underground as is at the surface (due to the proximity of rock walls). With shielding the background decreases by a factor 100 in the energy region mentioned \cite{CASPAR}. These two latter underground locations are vastly superior in terms of shielding against muons and neutrons, which reflects in reduced background in gamma-ray spectra at E$_{\gamma}>$2.7 MeV.

\section{Test case: the $^{13}$C+$^{12}$C reaction studies}
\par{The first reaction that we studied was $^{13}$C+ $^{12}$C, together with the group from IMP (Institute of Modern Physics) Lanzhou, China. This reaction leads to an activation appropriate for our tests:  $^{24}$Na, which has a half-life of 15 hours and is formed by one proton evaporation from the compound nucleus  $^{25}$Mg. Our choice of test case was motivated by the need  to test the characteristics of the facility as well as to study the fusion reaction mechanism deep under the Coulomb barrier in a system close to the $^{12}$C+$^{12}$C reaction of great importance in nuclear astrophysics. We studied the $^{13}$C+ $^{12}$C fusion reaction in the energy range of E$_{lab}$ =4.6 up to 11 MeV using the activation method and gamma-ray spectroscopy. That translates into an energy range E$_{cm}$=2.2 – 5.6 MeV, which is deep into the Gamow window for the  $^{12}$C+$^{12}$C burning at relevant stellar temperatures \cite{Iliadis}.
\begin{figure}[h]
	\begin{center}
		\includegraphics[width=0.48\textwidth]{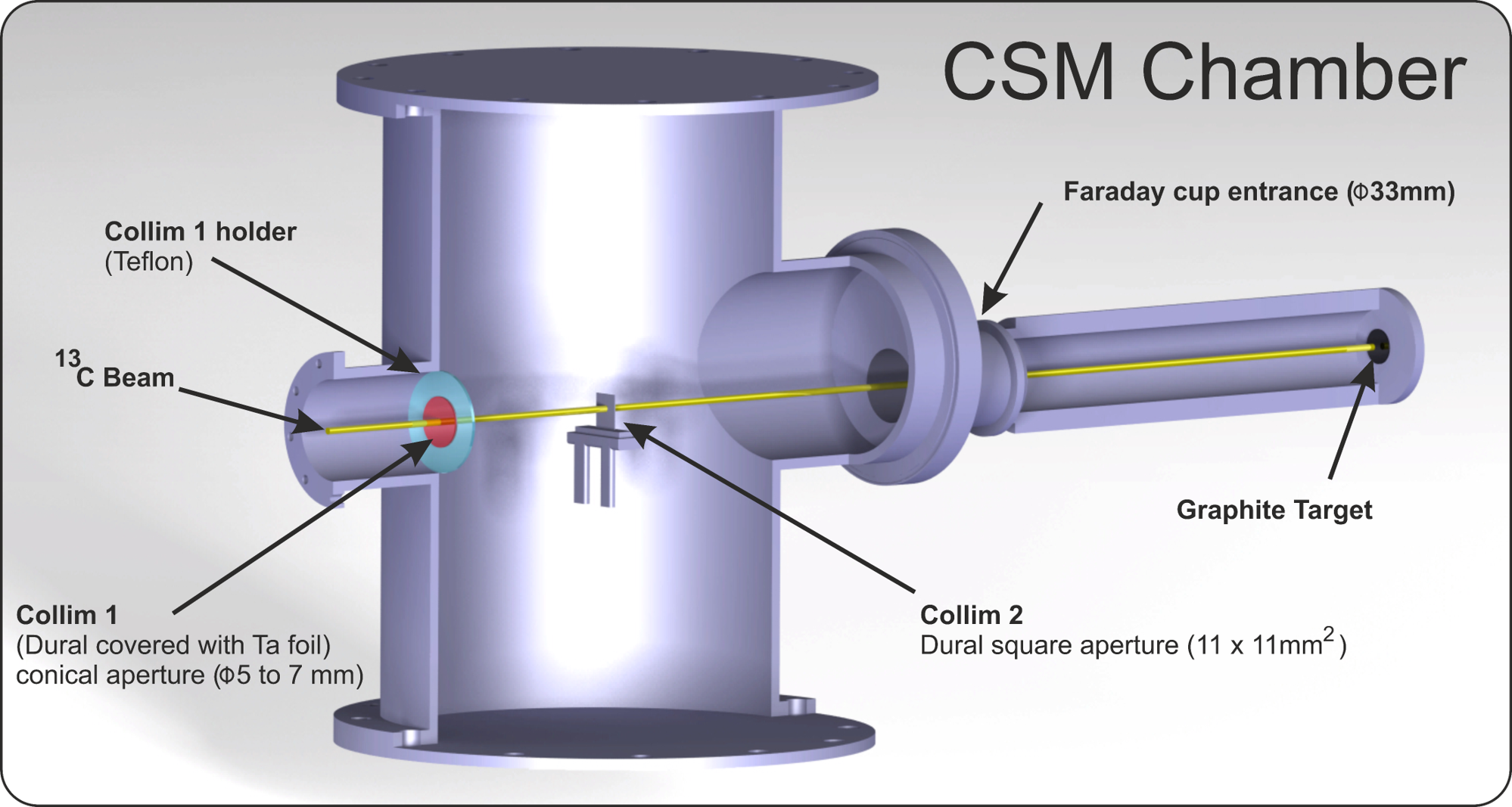}
	\end{center}
	\setlength\abovecaptionskip{-2pt}
	\caption{Irradiation chamber schematics.}\label{7}
\end{figure}	
Beams of $^{13}$C were obtained from the sputtering source with intensities of 0.4 up to 15 p$\mu$A and different charge states. The targets used were made of pure natural graphite with a thickness of 1 mm. For the energies where the irradiation time was longer it was necessary to cool-down the targets using a dielectric coolant system. After a number of tests, we found a situation where the current was reliably measured with the target that was also a Faraday cup (see Figure \ref{7}). A total of 71 targets were irradiated and measured.
For prompt gamma-ray measurements, a HPGe detector of 100\% relative efficiency was placed at 55$^{0}$ with respect to the beam axis in forward direction. We succeeded to determine the contributions from p, n and $\alpha$ evaporation channels for the energies where reaction cross sections were high enough to be measured in the accelerator hall (above 6.4 MeV). This is not a point important here, it is detailed further in Ref. \cite{Prompt}. The final irradiation setup is shown in Figure \ref{8}.
\begin{figure}[h]
\begin{center}
	\includegraphics[width=0.48\textwidth]{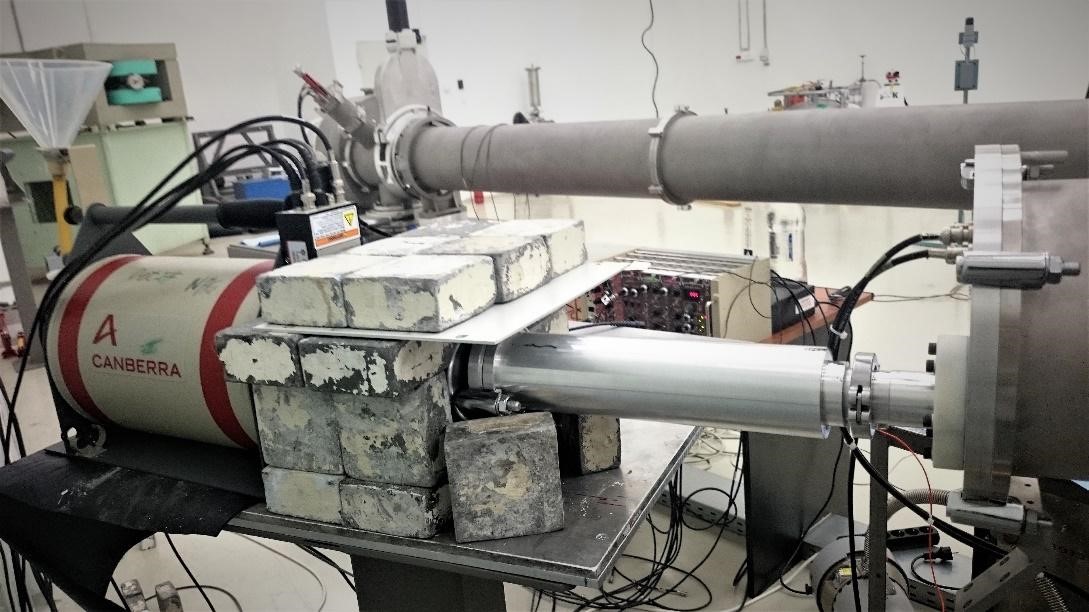}
	\end{center}
    \setlength\abovecaptionskip{-2pt}
	\caption{The irradiation chamber. A HPGe detector placed inside the lead castle to measure prompt gamma rays.}\label{8}
\end{figure}	
}
\par{The induced activities have been measured by detecting the $\gamma$ rays following the $\beta$ decay of $^{24}$Na reaction product with well shielded HPGe detectors at GammaSpec \cite{GammaSpec}, NAG (Nuclear Astrophysics Group - our group's laboratory) and $\mu$Bq laboratories. Because of its sufficiently long half time, $^{24}$Na was excellent for the procedure we used: up to one day of irradiation (depending on the incident energy), transfer to Slanic in 2.5 hours and about one day of de-activation measurements, during the irradiation of the next target, and so on. At bombarding energies higher than 5.6 MeV, activities were also measured in the certified setup GammaSpec situated next door, in a basement of the same department. In these 3 laboratories, the cascading gamma rays (1369 and 2754 keV) \cite{NNDC} were detected with HPGe detectors of 30\% (at GammaSpec), 100\% (at NAG) and 120\% (at $\mu$Bq in the salt mine) relative efficiency. For efficiency calibration we used sources with well-known activities, like: $^{152}$Eu, $^{133}$Ba, $^{60}$Co,$^{137}$Cs, $^{241}$Am.
	
The coincidence summing correction was determined by measuring one target placed in close (1 mm) and far geometry (15 cm). Calibrations and measurements performed in similar conditions allowed us to reduce the systematic uncertainties associated with the experimental data corresponding with the range E$ _{c.m} $ = 2.2-5.6 MeV below 10\%. Targets irradiated at the same energy and in similar conditions were measured at GammaSpec and in the salt mine. The differences between the results were within the estimated errors. As such we could verify the efficiency calibration of the 120\% relative efficiency HPGe detector used in the underground laboratory in Slănic and gives us confidence in the absolute values of the cross sections measured. Figure \ref{29} shows 3 spectra collected in the salt mine: (a) the gamma ray spectrum of a target that was irradiated at the energy of 8.6 MeV, (b) and (c) show the spectra of targets irradiated at E$ _{beam} $=4.8 MeV  without and with background subtraction. The background peaks left in (c) are due to variations in background during the 3.9 days of measurement of three different samples. These variations are very small and only visible relative to the very low activities of the samples we measured. They occur simply due to the presence/absence of the experimenters or of new devices in the measuring hall. The apparent  peak at E=2614 KeV is in fact due to statistical fluctuations in spectra subtraction of the very large numbers making the out-of-scale tall peak in (b).
\begin{figure}[h]
	\begin{center}
		\includegraphics[width=0.48\textwidth]{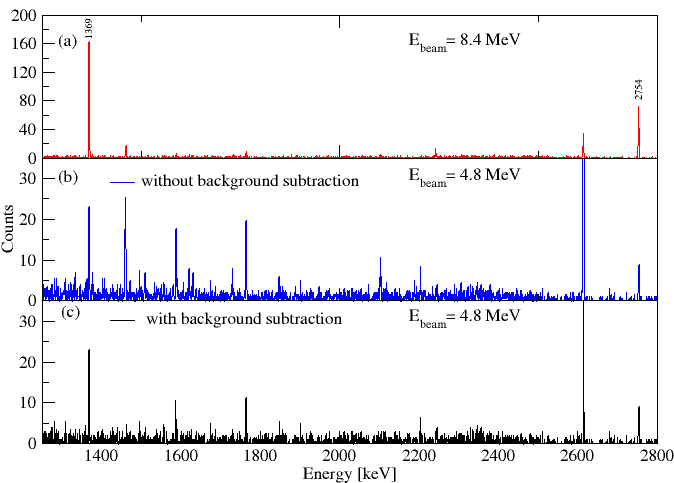}
	\end{center}
	\setlength\abovecaptionskip{-2pt}
	\caption{Spectra measured at $\mu$Bq for two targets irradiated at 8.4 (a) and 4.8 MeV (b, c). Only the two peaks of interest are labeled. The others are from background.}\label{29}
\end{figure}
}
\par{To determine the thick target yield we used only the 1369 keV peak which has a branching ratio of I$ _{\gamma} $=99.9935\% \cite{NNDC}. Firstly, the activity of the targets at the end of irradiation procedure and the beam current integrated in time (corrected step-wise for decay during irradiation) was determined:
\begin{equation}
\Lambda=\frac{\lambda C}{\varepsilon_{\gamma}I_{\gamma}t_{c}(1-e^{-\lambda t_{c}})}e^{\lambda\Delta t}
\end{equation}
where, C are the net counts of  full energy peak of a gamma transition, $\varepsilon_{\gamma}$ is the efficiency, I$ _{\gamma} $ the absolute branching ratio for 1369 keV $\gamma$-ray, t$_{c}$ is the counting time and $\Delta$t is the time between the end of irradiation and the start of counting.
\begin{figure}[h]
	\begin{center}
		\includegraphics[width=0.48\textwidth]{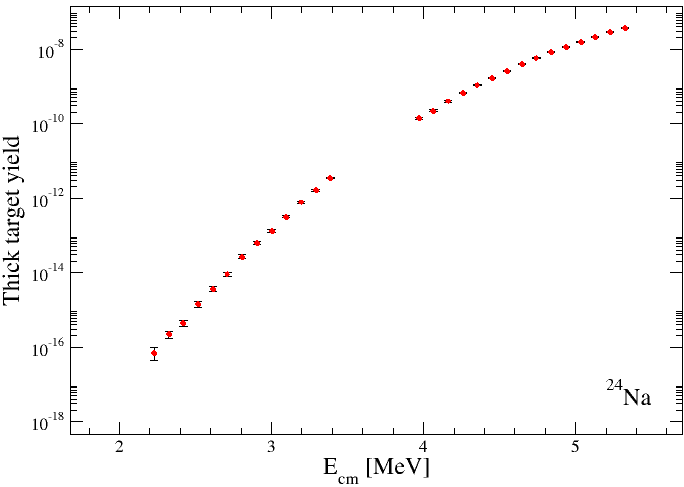}
	\end{center}
	\setlength\abovecaptionskip{-2pt}
	\caption{Thick target yield for the proton evaporation channel.}\label{9}
\end{figure}
Secondly, the thick target yield was determined as the ratio between the activity and beam current integrated in time (Fig. \ref10). The final step was to determine the proton cross section using the thick target method. 
Projectiles with different energies, in our case E and E-$\Delta$E, where $\Delta$E=0.2 MeV will penetrate two different depths, and the cross sections in Fig. \ref11 are determined by differentiating the yields and using stopping ranges calculated with SRIM \cite{SRIM}.
\begin{figure}[h]
	\begin{center}
		\includegraphics[width=0.48\textwidth]{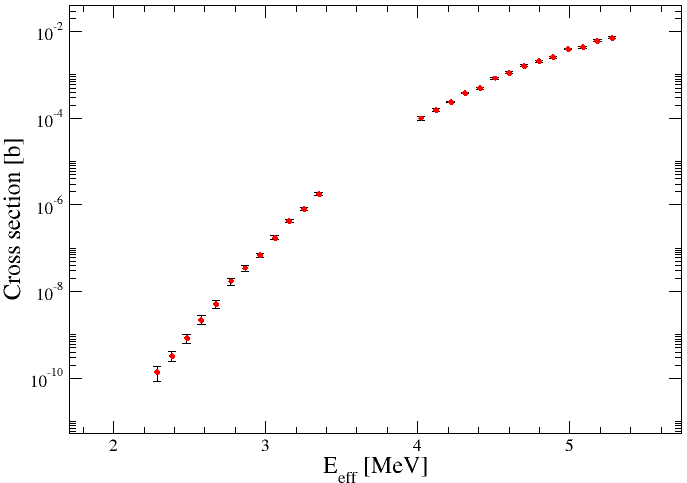}
	\end{center}
	\setlength\abovecaptionskip{-2pt}
	\caption{The proton evaporation, $^{12}$C($^{13}$C,p)$^{24}$Na, cross section from activation measurements for: E= 2.2-3.4 MeV and from E=4 up to 5.2 MeV, preliminary results.}\label{11}
\end{figure}
	
The activities of irradiated targets measured both in the underground and surface laboratories allowed to determine the limit of detection for cross sections to be of the order of ${\sim}$100 pb. Essentially, we achieved an increased  sensitivity of these measurements by about a factor 100 over the best experiments performed so far \cite{Notani}, which could only measure down to E$_{cm}$=2.64 MeV. Note that the latter used $\beta-\gamma$ coincidences to clean up the activation spectra.
\par{The facility was also used for the study of two other reactions induced by $\alpha$ particles. The first one was $\alpha$+$^{64}$Zn. During the experiment we irradiated natural thick zinc targets (1 mm) by alpha beams with energies in the laboratory frame between 5.4-8.0 MeV, in steps of 0.2 and 0.25 MeV. Total beam time was 140 h and the beam current varied from 0.2 to 0.7 $\mu$A. For this energy range we  measured the proton evaporation channel, $^{64}$Zn($\alpha$,p)$^{67}$Ga with a T$_{1/2}$=78.28 h, which is one of the three channels that lead to activation \cite{CSSP18}. In the underground laboratory gamma rays of E$_{\gamma}$=184.6, 209.0, 300.2 and 393.5 keV were detected \cite{NNDC}. In this case the sensitivity is increased again by a factor of around 100 compared with previous results \cite{Alpha}. For the second reaction, $\alpha$ beams impinged on natural Ni targets. In this case the reaction cross section for the ($\alpha$,$\gamma$) channel was determined with a similar sensitivity. The details of these experiments and the results will be the subject of other publications.

\section{Conclusions}
We present a new facility for nuclear astrophysics. We show that direct measurements can be successfully and reliably made using a small tandem accelerator for irradiations and an ultra-low background laboratory located underground in the Slănic-Prahova salt mine for de-activation measurements. Both facilities belong and are operated by IFIN-HH. We conclude and show that the accelerator is competitive to study reactions induced by alpha particles and light ions. After irradiation, the samples are transferred to the salt mine, about 120 km away from IFIN-HH Măgurele-Bucharest. The activity of the resulting samples is measured by high resolution, high efficiency HPGe detector(s) in an ultra-low gamma-ray background environment. This reduced natural radioactivity background is due to the natural conditions in the salt mine: the salt is pure with no radioactive contaminants, rocks are far away, and the salt walls are compact, with no cracks for radon gas to migrate. The salt mine is only about 210 m under surface (${\sim}$600 mwe), therefore cosmic radiation is not much inhibited, but its produced background is not important in our type of measurements. This procedure is limited for cases where the resulting activity has lifetimes larger than 1-2 hours, that is, comparable to the transfer time of the samples. If a gain of a factor around 100 in sensitivity is achieved - the actual number may differ depending on the de-activation gamma-rays, their energies and branchings - a loss of activity of 2 to 4 times during samples' transfer, may still allow an important gain in cases of real importance for NA. For shorter halftimes different methods will be applied to avoid the time consuming transfer of samples (primarily beta-gamma coincidences).
\par{Summarizing, the pros of this facility are:
	
- a small but stable accelerator with relatively high currents, including for light ions;

- a laboratory with ultra-low gamma-ray background in a salt mine for de-activation measurements with high resolution and efficiency HPGe detectors;

- the ability to determine absolute values for the cross sections measured - the GammaSpec utility is certified, cross calibrated internationally.

The limitation stems from the distance between the accelerator and the salt mine, which makes it inefficient for activities with half-lives shorter than 1-2 hours.}

 For the $ ^{13} $C+$ ^{12} $C case that was the most appropriate test of the procedure, activities as low as 3 mBq could be measured, and cross sections of about 100 pb. Two $\alpha$ induced cases were also studied, using the salt mine laboratory, with satisfying results. For example for the  $^{64}$Zn($\alpha$,p)$^{67}$Ga case we were able to determine a cross section of the order of 30 nb.

\textbf{Acknowledgments}
\\
This work was supported by Executive Unit for Financing Higher Education, Research, Development and Innovation (UEFISCDI), Bucharest, Romania under grant No. PN-III-P4-ID-PCE-2016--0743.
\bibliography{mybibfile}

\end{document}